\documentclass{aa}

\usepackage{fixltx2e}
\usepackage{graphicx}
\usepackage{txfonts}
\usepackage{multirow}
\begin{document}

\title{Properties of LBGs with [OIII] detection at z $\sim$ 3.5}
\subtitle{The importance of including nebular emission data in SED fitting}
\author{Fang-Ting Yuan\inst{1}, Denis Burgarella\inst{2}, David Corre\inst{2}, Veronique Buat\inst{2}, M\'{e}d\'{e}ric Boquien\inst{3}
 \and Shiyin Shen\inst{1}}

\institute{Shanghai Astronomical Observatory, CAS, Nandan Road 80, 200030, Shanghai, China \\
\email{yuanft@shao.ac.cn}
\and
Aix Marseille Universit\'{e}, CNRS, LAM (Laboratoire d'astrophysique de Marseille) UMR 7326, 13388, Marseille, France\\
\and
Centro de Astronom\'{i}a (CITEVA), Universidad de Antofagasta, Avenida Angamos 601, Antofagasta, Chile\\}

\date{Received xxx, 2019; accepted xxx}

\abstract
{
 Nebular emission lines are critical to measure physical properties in the ionized gas (e.g., metallicity, the star formation rate, or dust attenuation). They also account for a significant fraction of broadband fluxes, in particular at the highest redshifts, and therefore can strongly affect the determination of other physical properties, such as the stellar mass, which are crucial in shaping our understanding of galaxy formation and evolution.}
{We investigate a sample of 51 Lyman break galaxies (LBGs) at $3.0<z_{spec}<3.8$ with detected
[OIII] line emissions and estimated the physical properties of these galaxies to examine the impact of including nebular emission data in the Spectral energy distribution (SED) fitting.}
{We used the Code Investigating GALaxy Emission (CIGALE) to fit the rest-frame ultraviolet-to-near-infrared SEDs of these galaxies and their emission line data simultaneously. We ran CIGALE with and without the nebular model or the emission line data, and compare the results to show the importance of including the nebular emission line data in the SED fitting.}
{We find that without the nebular model, the SED fitting overestimates the stellar mass due to the strong [OIII] lines that are redshifted to the $K_s$-band, which is consistent with previous results. The emission line data are necessary to constrain the nebular model in the SED fitting. We examine the $K_s$-band excess, which is mostly used to estimate the emissions of the [OIII]$+$H$\beta$ lines when there is no spectral data, and we find that the estimation and observation are statistically consistent. However, the difference can reach up to more than 1 dex in some catastrophic cases, which shows the importance of obtaining spectroscopic measurements for these lines. We also estimate the equivalent width of the H$\beta$ absorption and find it negligible compared to the H$\beta$ emission.}
{Line emission is important to constrain the nebular models and to obtain reliable estimates of the physical properties of galaxies. These data should be taken into account in the SED fitting.}

\titlerunning{LBGs with [OIII] emission at $z\sim3.5$}
\authorrunning{Yuan et al.}

\maketitle

\section{Introduction} \label{sec:intro}

The Lyman break technique is commonly used to select high redshift galaxies. It was first used extensively by \citet{steidel1996} as the selection of U-band dropouts to detect galaxies at $z\sim 3$. The galaxies selected with the Lyman break technique, that is, Lyman Break Galaxies (LBGs), are found to be star forming galaxies with relatively low dust attenuation. As shown by \citet{madau2014}, these galaxies are important tracers for measuring the cosmic star formation rate (SFR) at $z>3$.

However, deriving the physical properties of LBGs can be very uncertain. First of all, the lack of spectroscopic redshift can lead to difficulties when estimating galaxy properties because the photometric redshift degenerates with properties, such as SFRs, age, metallicity, and reddening in the SED fitting. Second, at higher redshifts, the contribution of the nebular emission lines to broadband fluxes becomes greater than typically observed at intermediate redshifts. Third, the dust attenuation at high redshift is not well understood. The attenuation law is uncertain for high redshift galaxies \citep{kriek2013,buat2018}, and the data (especially infrared (IR) data) are not sufficient to constrain the dust attenuation. Furthermore, there are degeneracies between different parameters. For example, the age, metallicity, and dust attenuation degenerate with each other.

In this work, we focus on the effect of the nebular emission on the spectral energy distribution (SED) fitting. Nebular emission is found to be quite common in LBGs; \citet{debarros2014} show that about 65\% of LBGs present detectable signs of emission lines. These LBGs are more actively star forming and play an important role in estimating the cosmic SFR density as compared to the other 35\% of LBGs that have weak or no emission lines. However, the strong emission lines presented in these LBGs can contaminate the broadband photometry due to their larger equivalent width (EW). \citet{stark2013} show that [OIII]5007, H$\beta$, and $H\alpha$ contaminate $K$-band and {\it Spitzer}'s IRAC warm bands at 3.6$\mu$m and 4.5$\mu$m from redshift $\sim 3$. Therefore, it is essential to constrain emission lines by spectroscopic observation.

Spectroscopic emission line data not only enable us to remove the contamination from broadband photometry, but also give additional constraints for properties such as SFR, stellar mass, and dust attenuation. \citet{schaerer2013} suggest that observational emission line data may help constrain the star formation history (SFH). \citet{buat2018} show that including H$\alpha$ data in the fitting could provide constraints on the dust attenuation law.

Detecting emission lines at high redshift, however, is not trivial. It requires deep near infrared (NIR) spectroscopy with sufficient spectral resolution. When spectroscopic data are not available, emission lines are estimated from narrow band photometry or broadband SED fitting. A commonly used approach is to calculate the excess of the observed broadband fluxes relative to the SED modeled fluxes, assuming that the excess in $K$($K_s$) or IRAC bands is due to the line emission redshifted to the band \citep[e.g.,][]{stark2013,forrest2017,malkan2017}. This method provides a rough estimation for the emission lines, but its accuracy should be examined more carefully.

To examine the contamination from nebular emission lines and to constrain the properties of LBGs more accurately, we investigate a sample of LBGs with detected emission lines ([OII], [NeIII], H$\beta$, [OIII]4959,5007) at $3.0<z<3.8$. We analyze the properties of these galaxies using an SED fitting method by considering the data from both the broadband photometry and emission line observations. We attempt to determine the effect of including nebular emission data in the SED fitting, and whether and to what degree the nebular emission data can help constrain the physical parameters.

We introduce the data used in this work in Section \ref{sec:data} and describe the SED fitting method in Section \ref{sec:method}. In Section \ref{sec:result}, we present the fitting results and discuss the impact of including emission line data in the SED fitting. We further discuss the properties of the LBGs in our sample and the constraints of the data on the SED models in Section \ref{sec:discussion}. Finally, we conclude this work in Section \ref{sec:conclusion}.

\section{Data} \label{sec:data}

We aim to build a sample of LBGs that have both photometric and spectroscopic data. First, we collected a sample of LBGs with emission line data from the literature, including the following three works:

\citet[][hereafter S13]{schenker2013} present the fluxes of H$\beta$ and [OIII] lines for a sample of 20 galaxies at $z\sim3.5$ in the GOODS-North field. The NIR spectra were observed by Keck I MOSFIRE, and the fluxes were measured by fitting these lines with Gaussian profiles. For [OIII] doublet, only the total fluxes of [OIII]$\lambda4959$ and [OIII]$\lambda5007$ are presented.

\citet[][hereafter H16]{holden2016} present the fluxes of H$\beta$, [OIII]$\lambda4959$, and [OIII]$\lambda5007$ lines for 24 star forming galaxies at z $\sim$ 3.2{--}3.7 in the GOODS-South field. Similar to S13, these spectra are observed by MOSFIRE on the Keck I telescope. The line fluxes are extracted by fitting the Gaussian-Hermite model to the observed maps.

\citet[][hereafter T14]{troncoso2014} present the fluxes of [OIII]$\lambda5007$, H$\beta$, [OII]$\lambda3727$, and [NeIII]$\lambda3869$ lines for a sample of 34 galaxies. These galaxies were observed with the Assessing the Mass-Abundance redshift[Z] Evolution (AMAZE) and the Lyman-break galaxies Stellar population and Dynamics (LSD) project \citep{maiolino2008,mannucci2009}. The line fluxes were measured by fitting a Gaussian profile to the lines. The continua of the lines were estimated by linear interpolation and then subtracted from the spectra.

From the above selection, we obtain 78 galaxies that have emission line data. We note that these galaxies are shown to have little or no sign of AGN activities, so we ignore the effect of AGN in this work.

We then collected the photometric data of these galaxies from public available multiwavelength catalogs, including the GOODS-MUSIC catalog \citep[CDFS field,][]{grazian2006,santini2009A&A}, the 3D-HST catalog \citep[GOODS-S and GOODS-N field,][]{brammer2012,skelton2014}, and a catalog constructed by  \citet{magdis2010} (Table \ref{tab:sample}). In CDFS or GOODS-S field, a source may be found in both the GOODS-MUSIC catalog and the 3D-HST catalog. For such sources, we choose to use the measurements in the GOODS-MUSIC catalog, because it provides additional photometry for {\it Spitzer}'s MIPS 24$\mu$m band. We also compared the measured fluxes at each photometric band given by these two catalogs and confirm that the values agree when their uncertainties are considered. Therefore, which catalog to use does not affect our conclusion.

We matched the photometric data to the line data using a radius of 5\arcsec. The closest match was used if there are multiple matches. We also compared the redshifts given by the photometric catalog, $z_\mathrm{phot}$, with those from the line catalog, $z_\mathrm{line}$. The matches with $\Delta z/z_\mathrm{line}>0.1$ were removed from our sample, where $\Delta z$ is the difference between $z_\mathrm{phot}$ and $z_\mathrm{line}$.
In total, there are 51 galaxies whose matches are found in the photometric catalogs.

\begin{figure*}
\centering
\includegraphics[width=17.0cm]{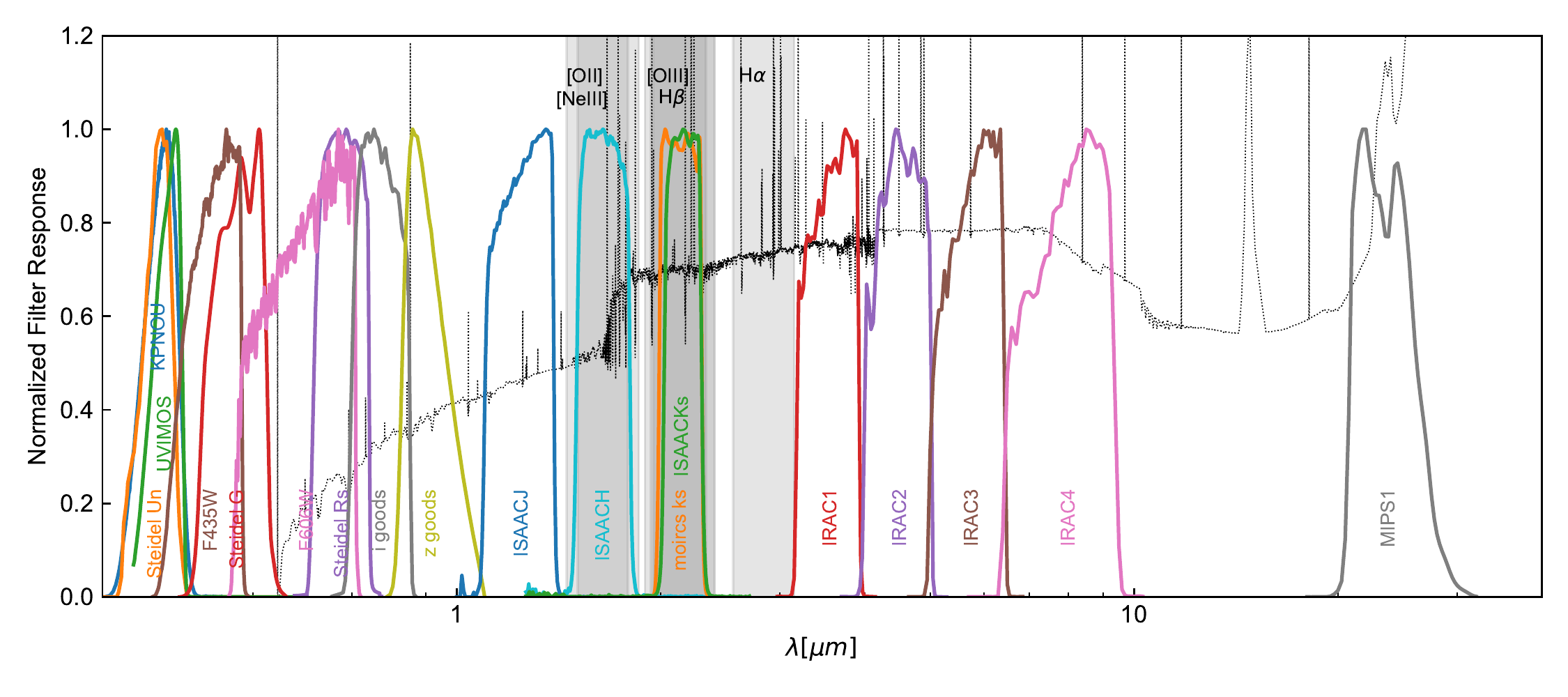}
\caption{Response of each filter used in this work. The shaded regions show the observed wavelengths of optical emission lines at $3.0<z<3.8$. The dotted line shows an example of the SED for a galaxy at z$\sim$3.5 in arbitrary scale.}
\label{fig:filters}
\end{figure*}

The photometric bands used in this work are summarized in Table \ref{tab:sample}. These bands were selected to obtain a good sampling of the SED from the ultraviolet (UV) to the IR. The wavelength and response of these bands are shown in Figure \ref{fig:filters}. To examine the contamination of optical emission lines, we plot in Figure \ref{fig:filters} the wavelength ranges of five main optical emission lines at redshift $3.0<z<3.8$. It can be observed that, at this redshift range, the $K_s$-band is contaminated by [OIII]$\lambda\lambda4959,5007$ and H$\beta$ lines. H-band is contaminated by [OII]$\lambda3727$ and [NeIII]$\lambda3869$ lines.

\begin{table*}
\centering
\caption{Photometric data of sample. \label{tab:sample}}
\resizebox{\linewidth}{!}{
\begin{tabular}{lcc}
\hline\hline
{Photometric catalog} & {Bands} & {Number of galaxies}\\
\hline
GOODS-MUSIC CDFS &  UVIMOS   F435W  F606W   i\_goods   z\_goods  ISAAC$J$, $H$, $K_s$   IRAC1{--}4, MIPS1 & 25\\
3D-HST GOODS-S & UVIMOS  F435W  F606W   i\_goods   z\_goods  ISAAC$J$, $H$, $K_s$   IRAC1{--}4  & 2\\
3D-HST GOODS-N & KPNOU   F435W   F606W    i\_goods    z\_goods F125W  F160W    MOIRCS\_$K_s$   IRAC1{--}4 & 15 \\
\citet{magdis2010} & Steidel\_Un, G, Rs    IRAC1{--}4 & 9\\
\hline
\end{tabular}
}
\end{table*}

Since the sample and data are collected from different works and catalogs, the quality of the data needs to be carefully controlled. The remaining 51 galaxies are further classified into ``Sample-A'' and ``Sample-B'' to better describe the quality of the data. Sample-A galaxies must satisfy the following criteria: First, the galaxy must be detected in at least five bands at an signal-to-noise ratio (S/N) higher than $3$; Secondly, the galaxy must be detected in at least one of the four IRAC bands at an S/N higher than $3$; Finally, the galaxy must have at least one emission line detection that has an S/N $>3$.

The galaxies not satisfying these criteria are classified into Sample-B. Galaxies in sample-A (26 galaxies) have better data quality than those in Sample-B (25 galaxies), and thus their properties should be better constrained by the SED fitting. Although galaxies in Sample-B have a poorer data quality than those in Sample-A, all these Sample-B galaxies are detected in at least three photometric bands. In both Sample-A and Sample-B, the upper limit data are included additionally. These upper limits are treated in the SED fitting using the method presented by \citet{sawicki2012} and \citet{boquien2019} (see also Section \ref{sec:method}).

\section{Method} \label{sec:method}

\begin{table*}
\centering
\caption{Parameters in SED fitting. \label{tab:sedpar}}
\begin{tabular}{cc}
\hline
SFH, delayed & \\
\hline
$\tau$ & 50, 100, 500, 1000, 2000.0, 5000, 10000\\
age & 10, 50, 100, 500, 1000, 2000\\
\hline
Stellar population, \citet{bc2003} & \\
\hline
IMF & \citet{salpeter1955} \\
metallicity & 0.0004, 0.004, 0.02 ($Z_{\odot}=0.02$) \\
\hline
Nebular, \citet{inoue2011} & \\
\hline
$f_\mathrm{esc}$ & 0.2 \\
$\log U$ & -2.5 \\
\hline
Dust attenuation, modified C00 & \\
\hline
$E(B-V)_\mathrm{lines}$ & 0.05, 0.1, 0.15, 0.2, 0.25, 0.3, 0.4, 0.5, 0.6, 0.7, 0.8, 0.9, 1.0, \\
 & 1.1, 1.2, 1.3, 1.4, 1.5\\
$\delta$ & -0.5, -0.4, -0.3, -0.2, -0.1, 0.0, 0.1, 0.2, 0.3, 0.4\\
$f_\mathrm{ebv}$ & 0.1, 0.2, 0.3, 0.4, 0.5, 0.6, 0.7, 0.8, 0.9, 1.0\\
Line emission extinction & MW extinction law for lines, Rv=3.1\\
\hline
Dust emission, \citet{dale2014} & \\
\hline
$\alpha$ & 2.0 \\
\hline
\end{tabular}
\end{table*}

\begin{figure*}
\centering
\includegraphics[width=0.99\textwidth]{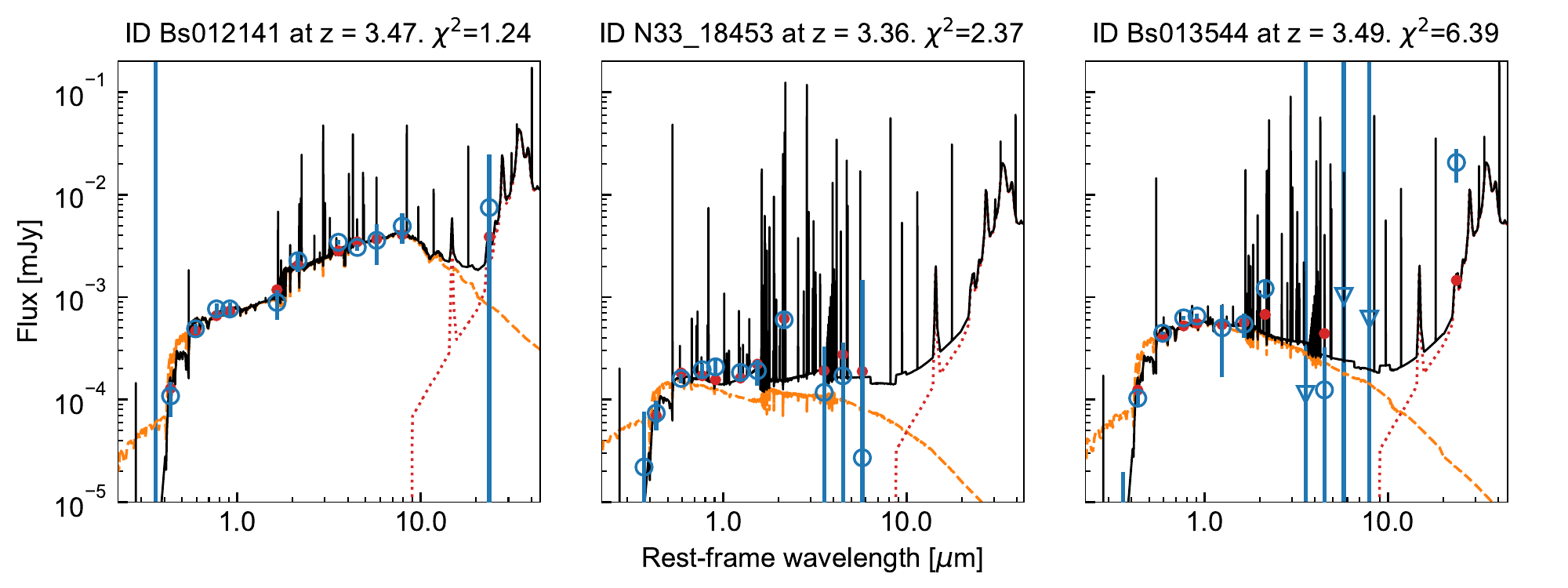}
\caption{Examples of SED fit with $\chi^2_r=1.24$ (\textit{Left}), $\chi^2_r=2.37$ (\textit{Middle}), and $\chi^2_r=6.39$ (\textit{Right}). The best model from CIGALE SED fitting is shown in solid line. The observed fluxes (open circles), the model spectrum of the stellar population (dashed line), the model fluxes (dots), and the dust emission (dotted line) are overplotted. For the fit with $\chi^2_r>5$ (\textit{Right}), the discrepancy between the observed fluxes and the best model is significant.}
\label{fig:fitqual}
\end{figure*}

\begin{table}
\centering
\caption{Models and Data used for the Withline, Noline and Nonebular runs of CIGALE carried out in this work. \label{tab:runs}}
\begin{tabular}{ccc}
\hline
Name  &  Models  &  Data \\
\hline
Withline &  stellar+nebular  &  photometric + line \\
Noline  &  stellar+nebular  & photometric \\
Nonebular & stellar  &  photometric \\
\hline
\end{tabular}
\end{table}

\begin{figure*}
\centering
\includegraphics[width=\linewidth]{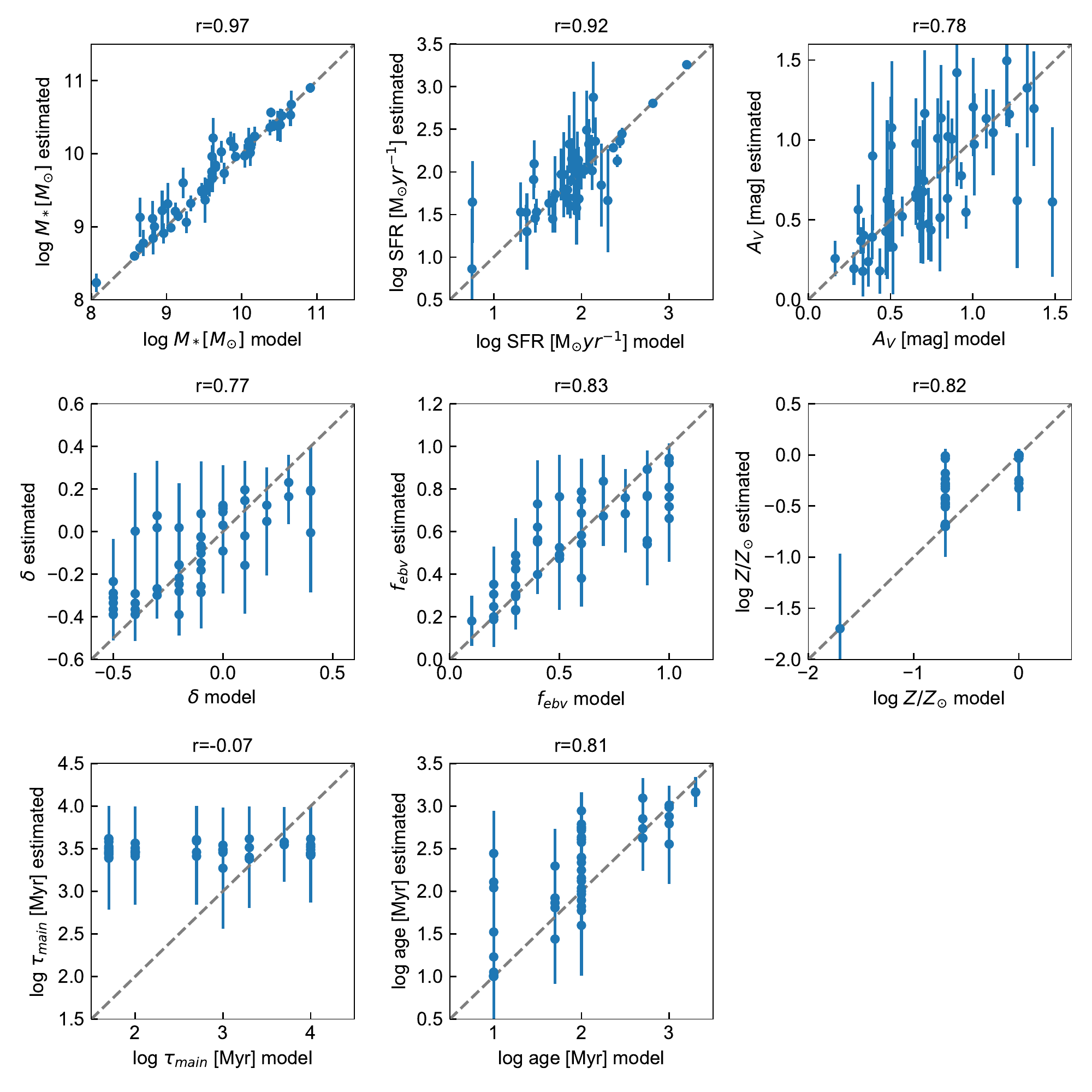}
\caption{Results of SED fitting of the mock catalog. Each plot corresponds to an output parameter considered in this work. The model values are plotted on the $x$-axis, the results applying the SED fitting method to the mock data are plotted on the $y$-axis, and the standard error given by the Bayesian analysis is over-plotted as an error bar for each value. The linear Pearson correlation coefficient ($r$) is indicated above each plot. \label{fig:mock}}
\end{figure*}

\subsection{SED modeling with CIGALE} \label{subsec:cigale}

The SED fitting is performed with Code Investigating GALaxy Emission (CIGALE)\footnote{https://cigale.lam.fr}. Detailed descriptions of the code CIGALE can be found in \citet{burgarella2005}, \citet{noll2009} and \citet{boquien2019}, as well as on the web site. This code adopts the energy balance strategy which assumes that the energy absorbed by dust is re-emitted at IR wavelengths so that it can fit the star and dust emission simultaneously. For this study, we make use of a particular feature of CIGALE that it can fit the photometric data and the nebular emission data simultaneously.

We use the \citet{salpeter1955} initial mass function (IMF) and the stellar populations from \citet{bc2003}. The SFH is assumed to follow the delayed form:
$SFR \propto t/\tau^2 \exp(-t/\tau)$, where $t$ is the age from the onset of star formation, and $\tau$ is the star formation time-scale. This functional form is more flexible and more physical than the simple exponentially increasing or decreasing SFH. It removes the discontinuity in SFR at $t=0$ and can mimic an increasing SFR when $\tau$ is large \citep{hirashita2017,carnall2019}. More discussion of the SFH models is presented in Section \ref{sec:discussion}.

We assume that the dust attenuation for the stellar part obeys the modified Calzetti Law \citep{noll2009}, which modifies the original Calzetti attenuation curve \citep[][hereafter C00]{calzetti2000} by adding a UV bump and by adjusting the slope. In this work, we adopt the form of the attenuation law assuming there is no bump:
\begin{equation}
    A(\lambda)=E(B-V)_\mathrm{star}k'(\lambda)\left(\frac{\lambda}{\lambda_\mathrm{V}}\right)^\delta,
\end{equation}
where $k'(\lambda)$ is from C00, and $\delta$ indicates the modification to the slope. The input value of $\delta$ is taken as free, in the range of -0.6 to 0.4. We assume there is no bump because our broadband data are unable to constrain the bump. Further observation with better spectral resolution are required to provide a better constraint.

The nebular emission is modeled based on the templates of \citet{inoue2011} using the CLOUDY code. The input parameters include the metallicity, radiation strength $U$, the dust attenuation, and the escape fraction of the Lyman continuum photons, $f_\mathrm{esc}$. The metallicity is assumed to be the same as that of stars. The radiation strength $U$ is fixed to $\log U=-2.5$. The dust obscuration for the nebular emission is assumed to follow a simple screen model and a Milky Way extinction curve \citep{cardelli1989}. Since our emission line data are at optical wavelengths, other ``standard'' extinction curves (e.g., LMC, SMC) produce similar extinction values \citep{fitzpatrick1986,calzetti1997}. The $f_\mathrm{esc}$ is fixed to 0.2 \citep[e.g.,][]{stark2013}. In our SED fitting, we estimate the model fluxes of the emission lines using a linear interpolated continuum obtained from the synthetic model spectrum in order to be consistent with what has been measured from the observations. CIGALE also allows to use the theoretical fluxes of the emission lines (given by CLOUDY) as the model fluxes. We find little difference between the results of these two methods.

The nebular emission and stellar emission are connected through $f_\mathrm{esc}$ and $f_\mathrm{ebv}$. With the escape fraction $f_\mathrm{esc}$, there are $(1-f_\mathrm{esc})$ of Lyman continuum photons to ionize the gas and power the nebular emission. The total number of Lyman continuum photons, which determines the absolute strength of both the continuous and line emission of nebula, is computed from the stellar SED \citep[e.g.,][]{sdb2009,castellano2014}.
The parameter $f_\mathrm{ebv}$ is the ratio of the $E(B-V)_\mathrm{star}$ to the $E(B-V)_\mathrm{gas}$. Although $f_\mathrm{ebv}=0.44$ given by \cite{calzetti1997} is commonly adopted in the works on the nearby galaxies, many recent studies have found that this factor can
vary up to $1$ \citep[e.g.,][]{puglisi2016,yuan2018,buat2018}. In this work, we take this parameter as free and find that the fitting is improved compared to that using a fixed value.

CIGALE is also able to fit the dust emission. However, the SED shape of dust emission is not constrained in this study because of the lack of rest-frame far-IR data. We model the dust emission using the templates from \citet{dale2014} (without AGN component) with a single parameter $\alpha$ that represents the exponent of the distribution of the heating intensities over dust masses. We adopt the SED template with $\alpha=2.0$.
The input values of the model parameters are listed in Table \ref{tab:sedpar}.

The output parameters are estimated through a Bayesian approach. For each model, the $\chi^2$ is calculated by comparing the model fluxes and the observed fluxes. Including the nebular emission data, the $\chi^2$ is thus the sum of $\chi^2_\mathrm{lines}$ for emission lines and $\chi^2_\mathrm{pho}$ for photometric fluxes. For upper limits, the error function is used to calculate the $\chi^2$ (See Eq. (15) in \citet{boquien2019}). The probability distribution function (PDF) of each parameter is then calculated from the likelihood, which is proportional to $\exp(-\chi^2)$. The parameters and the corresponding uncertainties are thus estimated from the PDF by taking the likelihood-weighted mean and the standard deviation of all models.

The global quality of the fit is estimated using a reduced $\chi^2$ of the best model (that is, the model with the smallest $\chi^2$), $\chi^2_r=\chi^2/(N-1)$, where $N$ is the number of input fluxes. In this fitting, only one galaxy has $\chi^2_r>5.0$. The galaxy has large measurement errors in the IRAC bands and an abnormal shape of SED (Figure 2, right panel). It is probably mismatched between the IR bands and the optical bands in the photometric catalog. Therefore, we removed it from our sample. Figure \ref{fig:fitqual} shows three examples of the SED fitting with different value of $\chi^2_r$. For the fit with $\chi^2>5$, which was already removed from our sample, the best model deviates significantly from the observation.

\subsection{Reliability of the output parameters \label{subsec:reliability}}

We test the reliability of the parameters using the method developed by \citet{giovannoli2011} and widely used among CIGALE users \citep[e.g.,][]{buat2012, boquien2012,yuan2018}. The mock catalog is based on the fitting described in Section \ref{subsec:cigale}. The mock spectra of each galaxy are created from the best-fit model. The fluxes of the photometric bands are then generated by deviating the flux of the best-fit model in each band with a random error $\sigma$, which follows a Gaussian distribution with a standard deviation of the observed error in that band. On these simulated data, we run the SED fitting code and compare the ``accurate'' values known from the best-fit model and the values estimated by the code.

Figure \ref{fig:mock} shows the comparison result. Values of the linear Pearson correlation coefficient ($r$) are also indicated in the plots. We find very good correlations for $M_*$ and SFR with $r>0.9$. The parameters $f_\mathrm{ebv}$, age, and metallicity $Z$ are also well constrained ($r>0.8$). The estimations of $V$-band attenuation $A_V$ and $\delta$ are less satisfying but still marginally constrained, with $0.7<r<0.8$. The parameter $\tau_\mathrm{main}$  is not constrained at all, with $r\sim0$.

\begin{figure*}
    \centering
    \includegraphics[width=16cm]{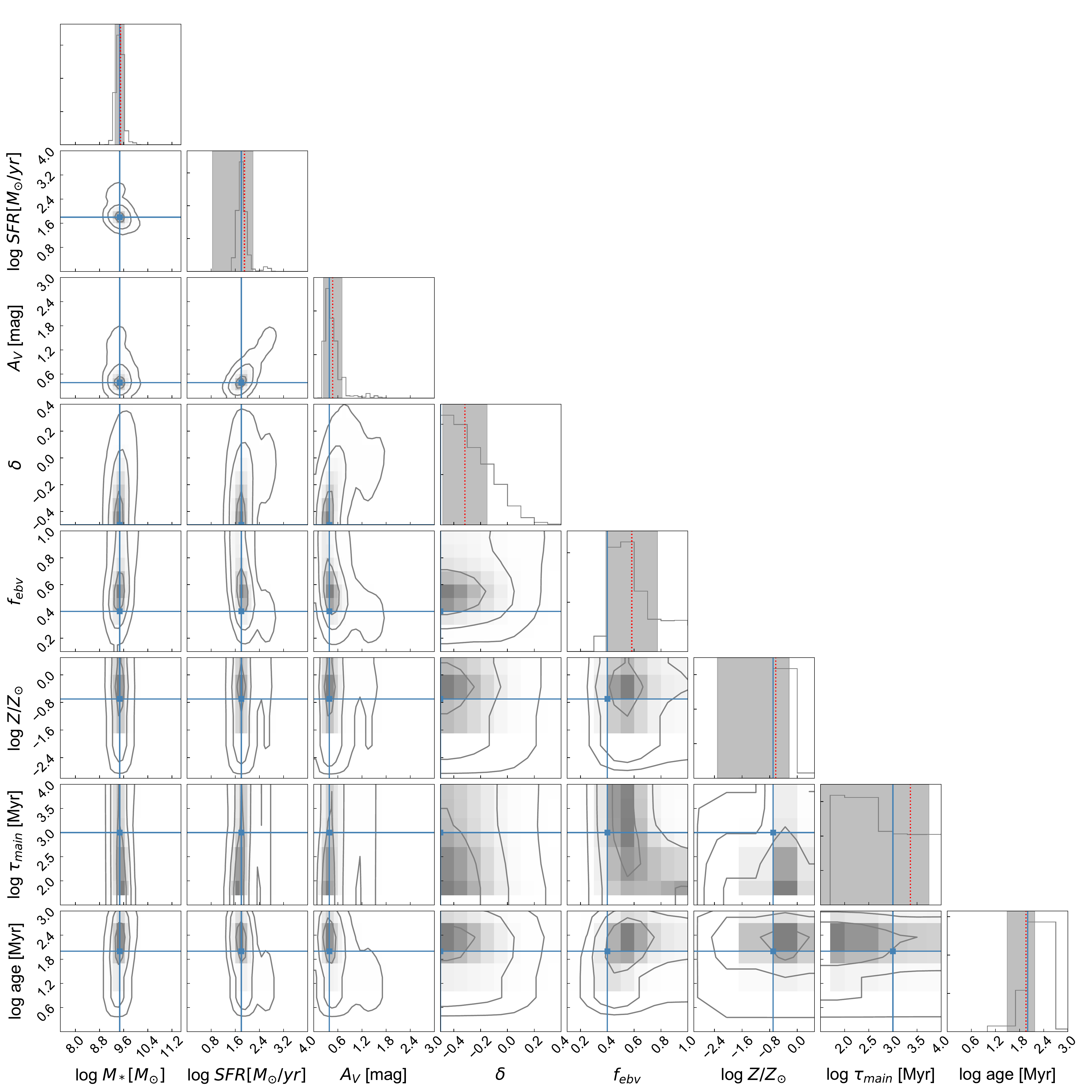}
    \caption{PDF corner plot for a mock galaxy with $\chi^2=1.34$, showing the 1D and 2D marginalized posteriors for each parameters using our Bayesian approach. The blue lines represent the true values used to generate the mock photometry. For the 1D histogram, the red dashed lines show the estimated values and the shaded regions indicate 1$\sigma$ uncertainties. For the 2D plot, the different contours correspond to 1, 2, 3$\sigma$.}
    \label{fig:cornerplot}
\end{figure*}

The Bayesian approach allows one to transparently test how well the data constrain certain parameters according to the shape of the resulting PDF. If a parameter is poorly constrained, we will obtain PDFs with large width which give large errors for the estimation \citep{salim2007}. The potential dependency and degeneracy of the parameters can be examined using the 2D PDF for each object. We show in Figure \ref{fig:cornerplot} an example of the 1D and 2D PDF of one mock galaxy with $\chi^2_r=1.34$, slightly above the average $\chi^2_r$ for our sample, which is $\sim$1.25. The figure shows that the model values are consistent with the estimated values considering the uncertainties. No severe degeneracy is found between these parameters. We also find that $\tau_\mathrm{main}$ is not constrained given the almost flat PDF, meaning that the probability for each value of $\tau_\mathrm{main}$ is the same. The estimated $\tau_\mathrm{main}$ thus depends on the input values. The 2D PDFs also show that although the parameters are estimated independently from each other, the global set of parameters (except the unconstrained ones) correspond to a self-consistent realization of the models, because the estimated parameters are all taken from models with high likelihood.

According to the above analysis which was also confirmed by that of the Bayesian error distribution \citet{noll2009}, we conclude that the reliable parameters are $M_*$, SFR, $Z$, $f_\mathrm{ebv}$ and age. In the following, we mainly use these parameters and also give some discussion about $A_V$, and $\delta$. We ignore $\tau_\mathrm{main}$ since it cannot be constrained.

\subsection{Fitting to the SEDs} \label{subsec:sedfit}

\begin{figure*}
\centering
\includegraphics[width=15cm]{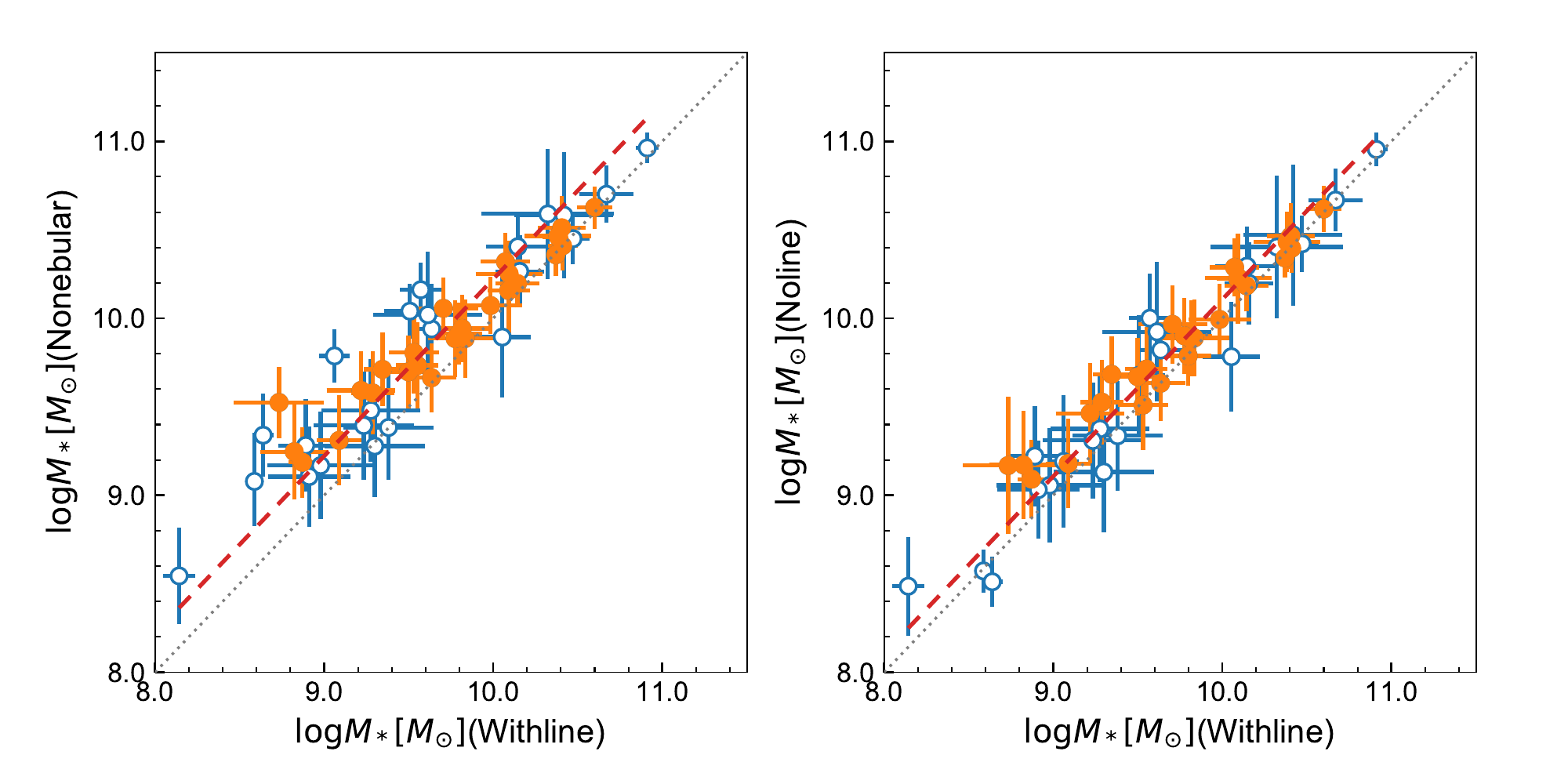}
\caption{\textit{Left}: Comparison of stellar masses derived from Withline and Nonebular runs. \textit{Right}: Comparison of the stellar masses derived from the Withline and Noline runs. Orange dots are galaxies in Sample-A. Blue circles are galaxies in Sample-B. Sample-A has better data quality than Sample-B, as defined in Section \ref{subsec:sedfit}. The dotted line is the 1:1 line. The dashed line shows the average difference between the two runs in each panel. \label{fig:stellarmass}}
\end{figure*}

First, we fit the SED using both the stellar and nebular models on both the photometric and line data. The models and input parameters for this fit have been described in Section \ref{subsec:cigale}, and we refer it as the ``Withline'' fit. In order to investigate the impact of the emission lines on the SED fitting, we additionally run two other fits, referred to as the ``Noline'' and ``Nonebular'' fits, respectively. In the Noline run, we use both the stellar and nebular models, which are exactly the same as in the Withline run, but only fit the photometric data. In the Nonebular run, we use only the stellar model, and fit for the photometric data.
The three fits are summarized in Table \ref{tab:runs}. We discuss the influence of including nebular emission model and data in the fitting by comparing the results of these three runs.

\section{Results}\label{sec:result}

\begin{figure*}
\centering
\includegraphics[width=15cm]{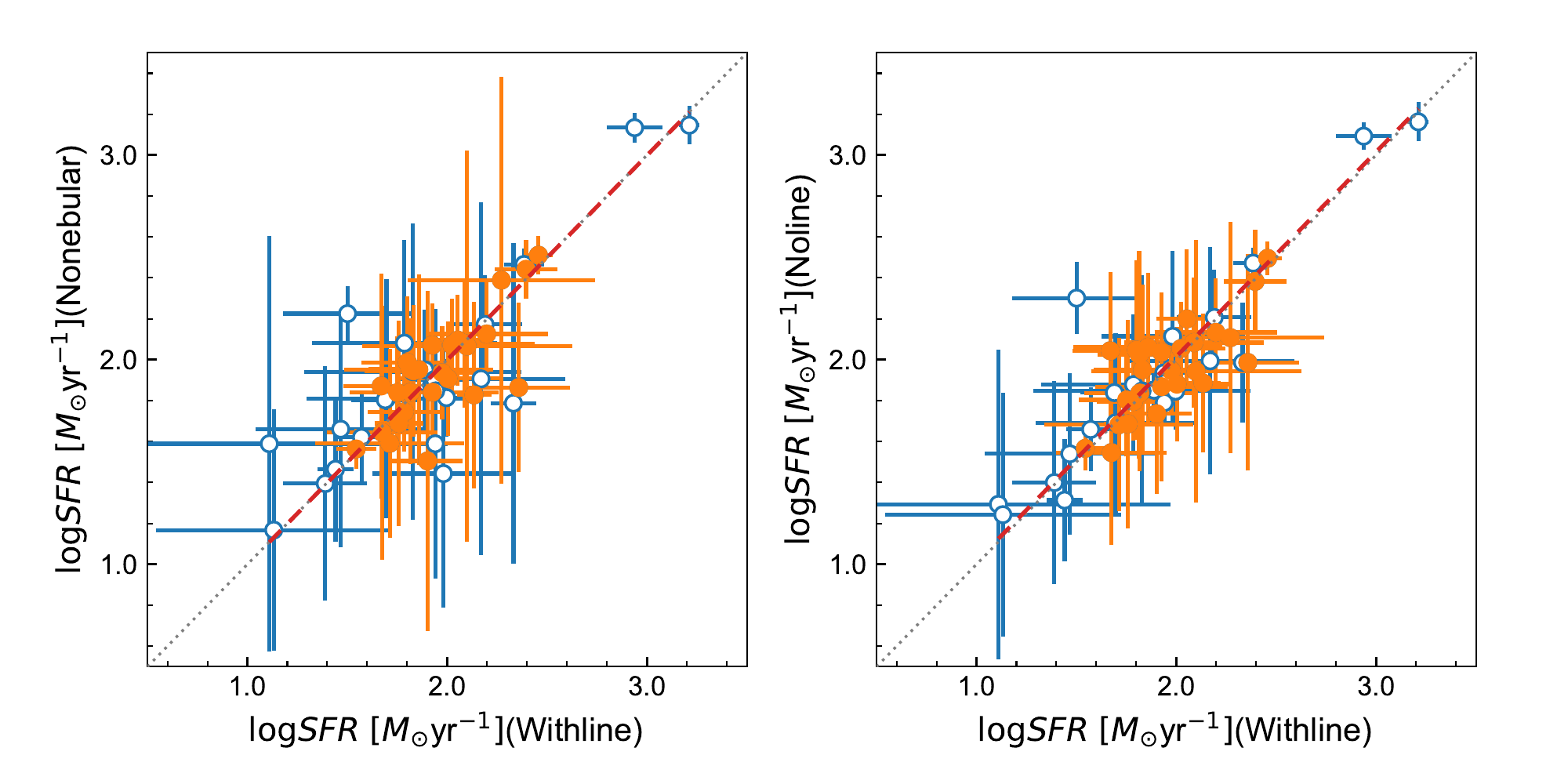}
\caption{\textit{Left}: Comparison of the SFRs derived from the Withline and Nonebular runs. \textit{Right}: Comparison of the SFRs derived from the Withline and Noline runs. Orange dots are galaxies in Sample-A. Blue circles are galaxies in Sample-B. Sample-A has better data quality than Sample-B, as defined in Section \ref{subsec:sedfit}. The dotted line is the 1:1 line. The dashed line shows the average difference between the two runs in each panel. \label{fig:sfr}}
\end{figure*}

\begin{figure*}
\centering
\includegraphics[width=15cm]{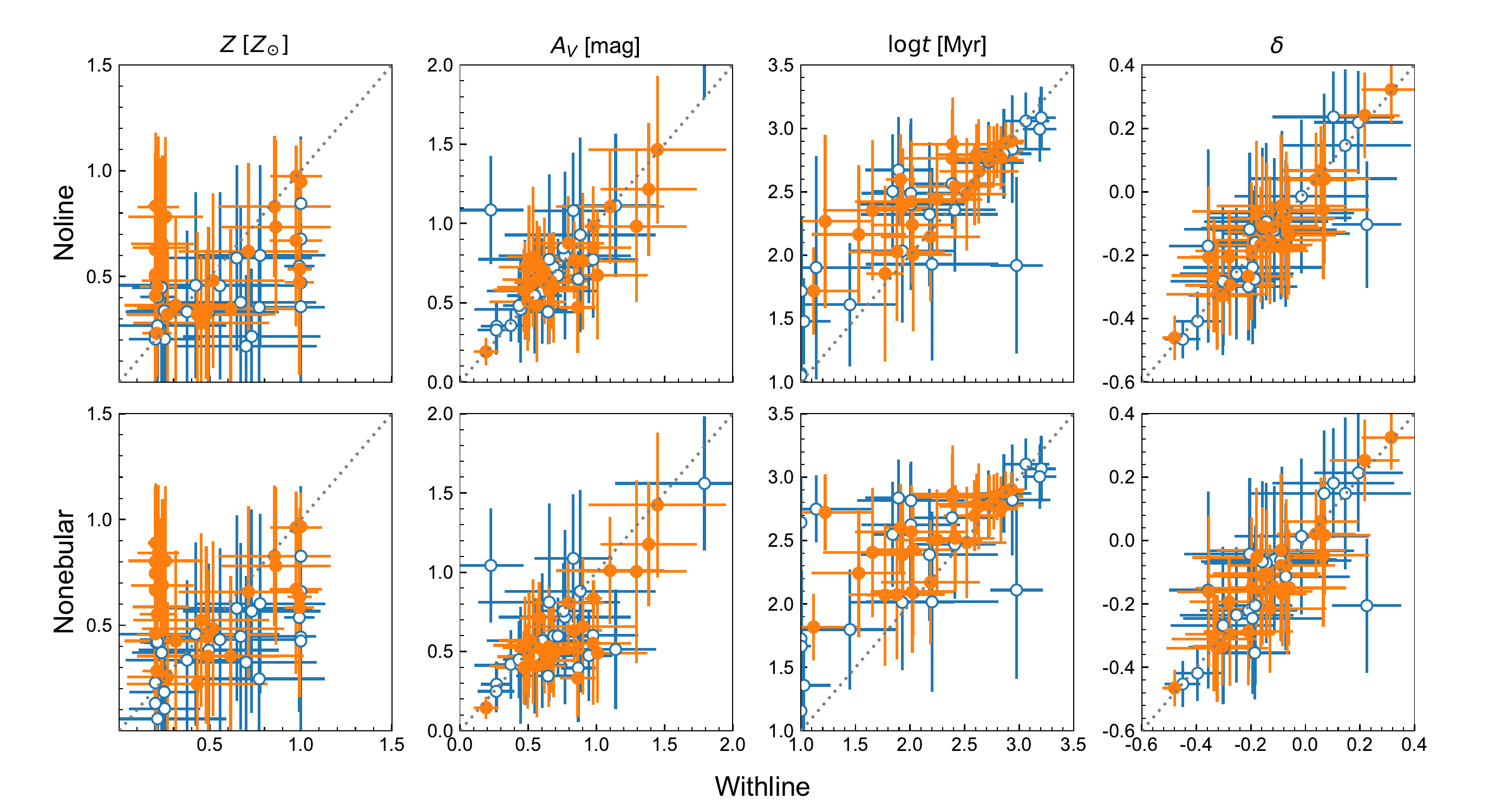}
\caption{\textit{Top}: Comparison of the metallicity, dust attenuation, age, and $\delta$ derived from the Withline and Noline runs. \textit{Bottom}: Comparison of the metallicity, dust attenuation, age, and $\delta$ derived from the Withline and Nonebular runs. Orange dots are galaxies in Sample-A. Blue circles are galaxies in Sample-B. Sample-A has a better data quality than Sample-B, as defined in Section \ref{subsec:sedfit}. The dotted line are the 1:1 line. \label{fig:4prop}}
\end{figure*}

\subsection{Impact of including line emission in SED fitting on the output parameters}\label{subsec:compprop}

Figure \ref{fig:stellarmass} compares the stellar masses derived from the Withline, Noline and Nonebular runs. We use the Withline run as a reference because this run included more data than the other two runs. Then we compare the other two runs to this reference run. First, compared with the Withline run, the Nonebular run overestimates the stellar mass by 0.22 dex, with a scatter of 0.21 dex. In the Noline run, with the nebular model included in the SED fitting, the consistency with the Withline run is improved: The average discrepancy is reduced to 0.11 dex, with a scatter of 0.14 dex. The difference between the Nonebular and Noline estimated stellar masses are 0.11 dex on average. This is consistent with the results of \citet{grazian2015}, who found that the stellar masses computed with and without the nebular contribution are less than 0.1-0.2 dex. For Sample-A and Sample-B, the conclusion seems quite similar, implying that the data quality affects insignificantly in estimating the stellar masses.

We then compare the SFRs derived from the three runs (Figure \ref{fig:sfr}). The Nonebular run gives very similar SFRs compared with the Withline run. The average difference is only 0.01 dex, with a scatter of 0.22 dex. The average difference between the Noline run and Withline run is also small, about 0.02 dex, with a scatter of 0.18 dex. It seems that the SFRs derived from the SED fitting are not affected significantly by including either the nebular model or the emission line data. The possible reason is that since we do not have H$\alpha$ data, the SFR derived from the SED fitting is heavily determined by UV fluxes, which has little contamination by emission lines. We also find that for Sample-A and Sample-B, the average difference between these runs is quite similar. The scatter of Sample-A is slightly smaller (by 0.04 dex) than Sample-B, for both the Withline-Nonebular and the Withline- Noline comparison.

We also compare the metallicity $Z$, the dust attenuation $A_{V}$, the age of the stellar population $t$, and $\delta$ derived from these three runs, as shown in Figure \ref{fig:4prop}. The Noline fit provides similar results compared with the Withline runs for these parameters. It slightly underestimates the metallicity and overestimates the age.
The Nonebular run, however, obtains lower metallicity, lower dust attenuation, and older age compared with the Withline run, for both Sample-A and Sample-B galaxies. For the modification of the C00 slope $\delta$, the three runs show little difference. It is not surprising because $\delta$ is not well constrained. As shown in Section \ref{subsec:reliability}, the correlation between the model values and the estimation of $\delta$ is only 0.77 for the Withline fit.

The different outputs of these three runs can be explained by the effect of the $K_s$-band excess caused by the strong line emission. In the Nonebular run, without any consideration of the nebular emission, the $K_s$-band fluxes are fully attributed to the stellar emission, and therefore the fit overestimates the amount of old stellar populations. The older ages then leads to a redder slope of the unobscured stellar models and to lower $A_V$ derived from the fit. The result is consistent with what was found by \citet{debarros2014} for galaxies with strong emission lines.

By adding the nebular model, the Noline run improves the results. However, due to the lack of constraints on the nebular models, the Noline fit still has some discrepancies with the Withline fit because in the Withline fit the emission line data provide additional constraints on the parameters.

\subsection{Impact on the parameter uncertainties}

\begin{figure}
    \centering
    \includegraphics[width=\linewidth]{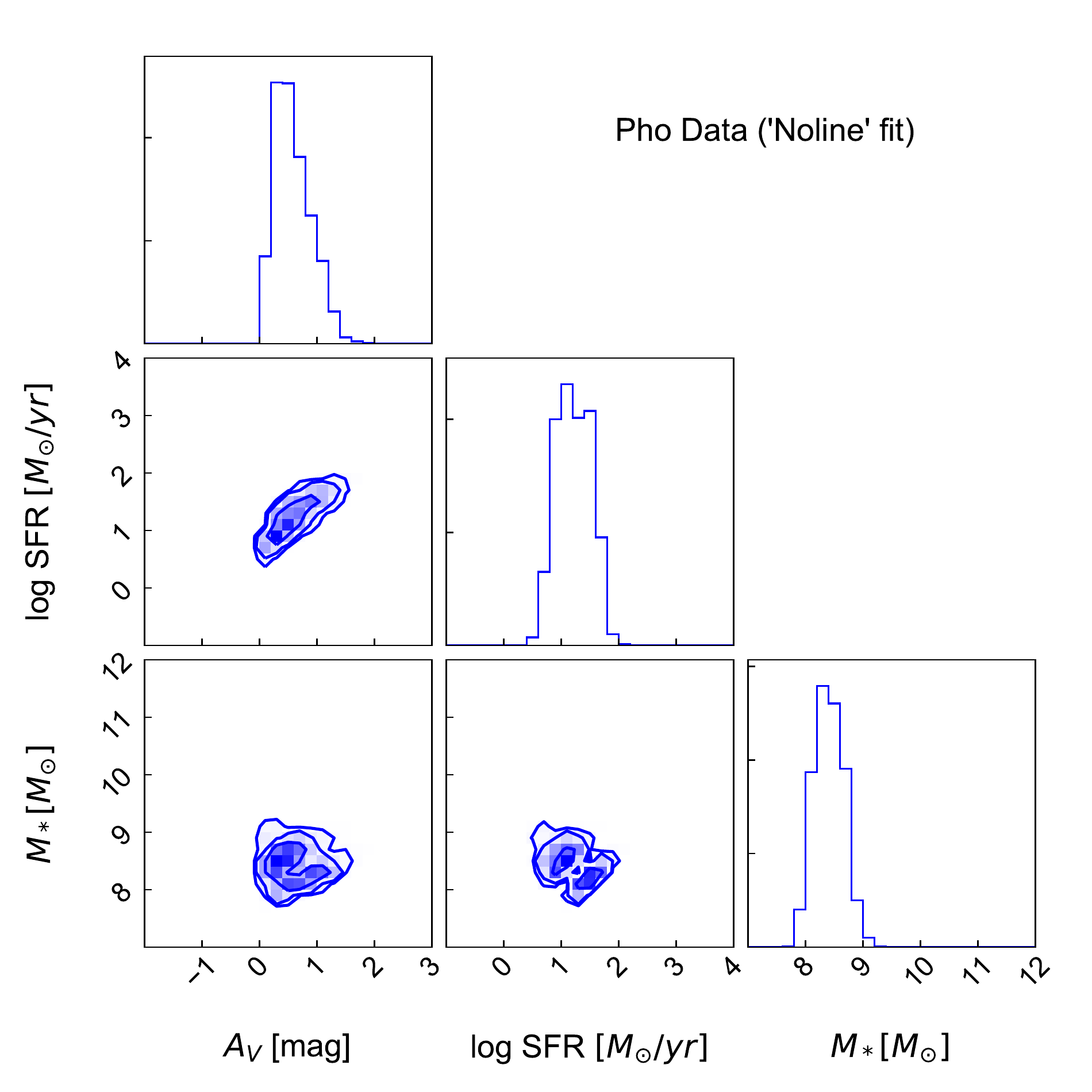}
    \includegraphics[width=\linewidth]{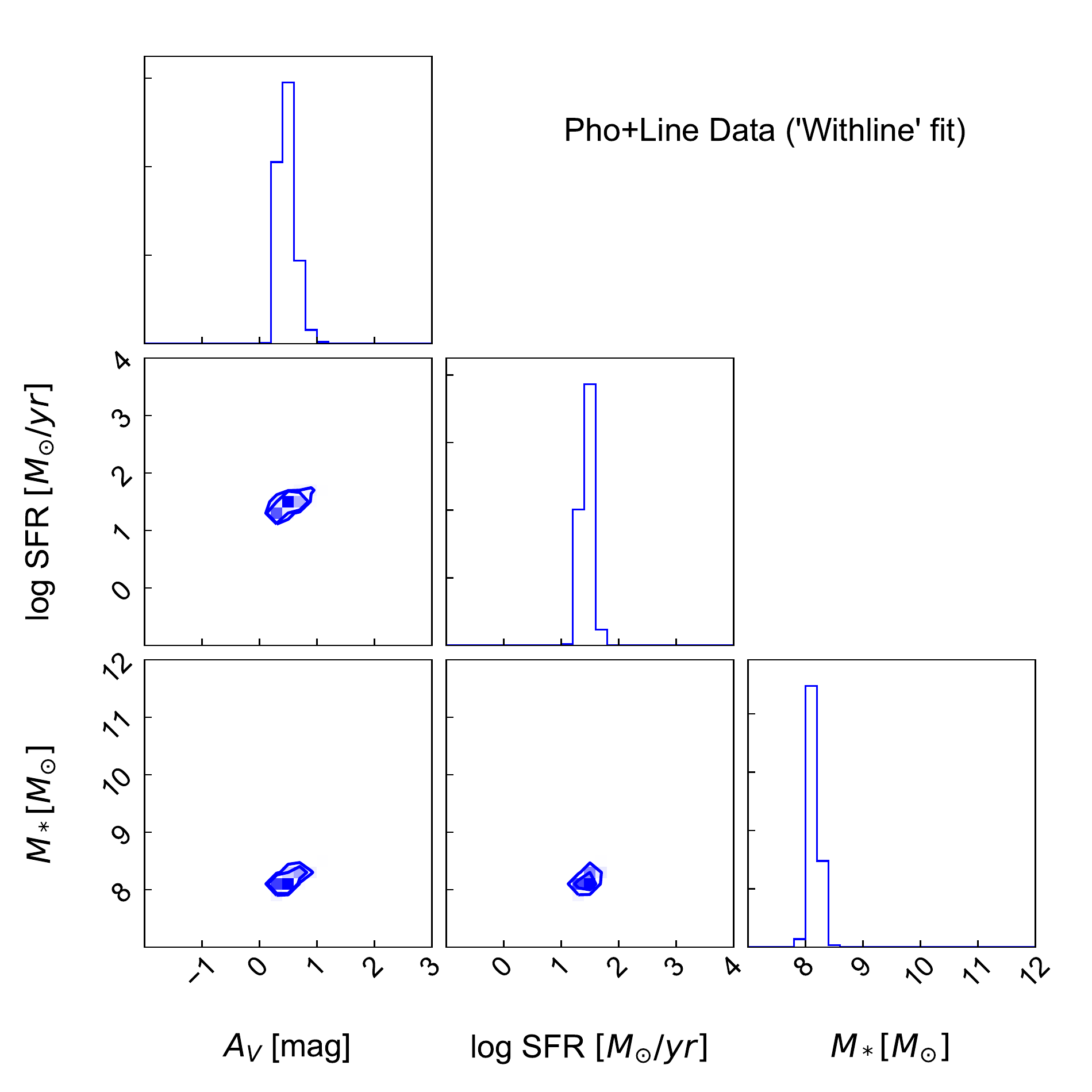}
    \caption{PDFs of $M*$, SFR and $A_V$ for a galaxy in our sample. Top: PDFs estimated from the Noline fit. Bottom: PDFs estimated from the Withline fit.}
    \label{fig:constraint}
\end{figure}

\begin{figure*}
\centering
\includegraphics[width=15cm]{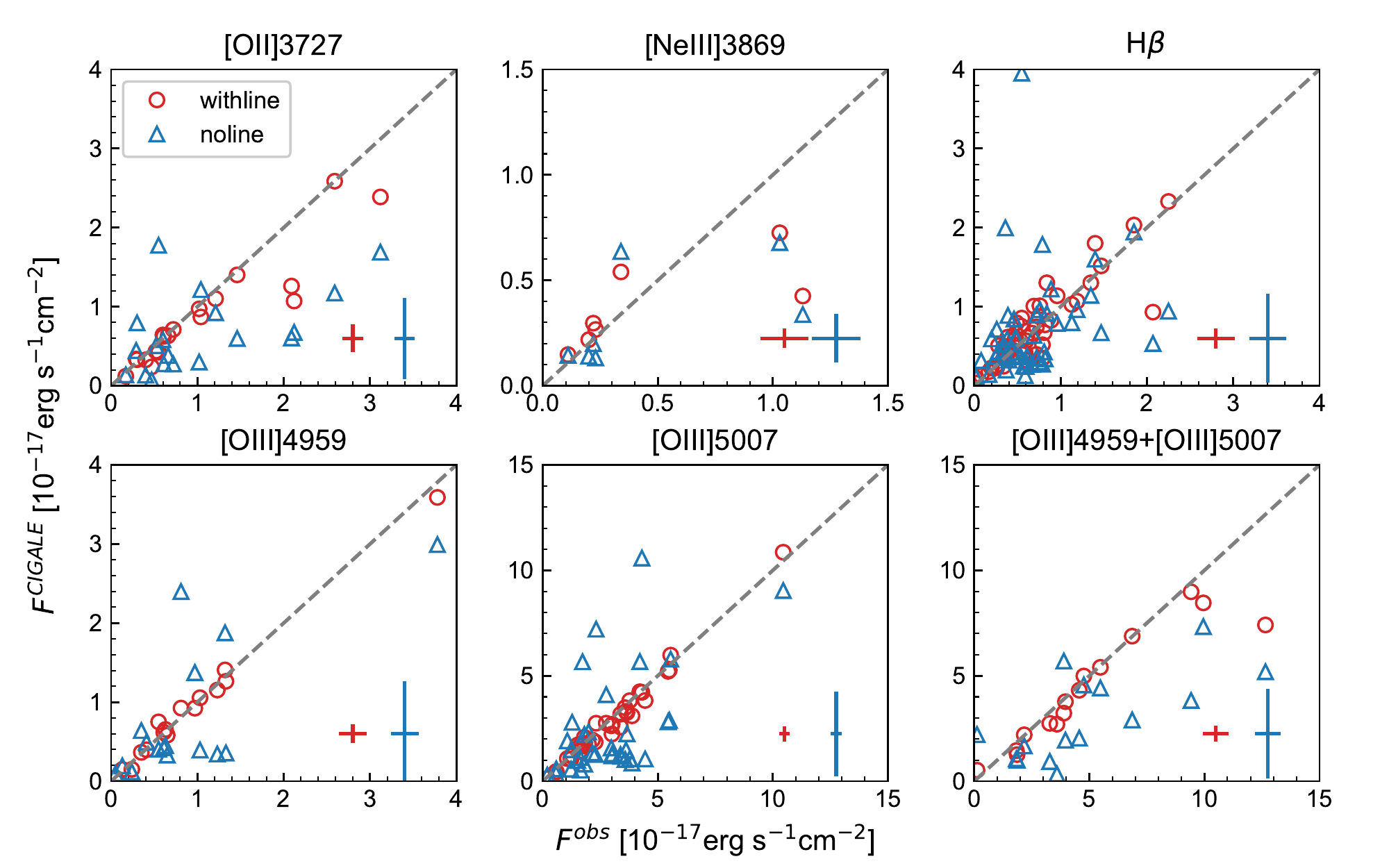}
\caption{Comparison of the observed and CIGALE modeled line fluxes. Circles show the fluxes obtained from the Withline run. Triangles show the fluxes obtained from the Noline run.} \label{fig:linecomp_woline}
\end{figure*}

\begin{figure}
\centering
\includegraphics[width=\linewidth]{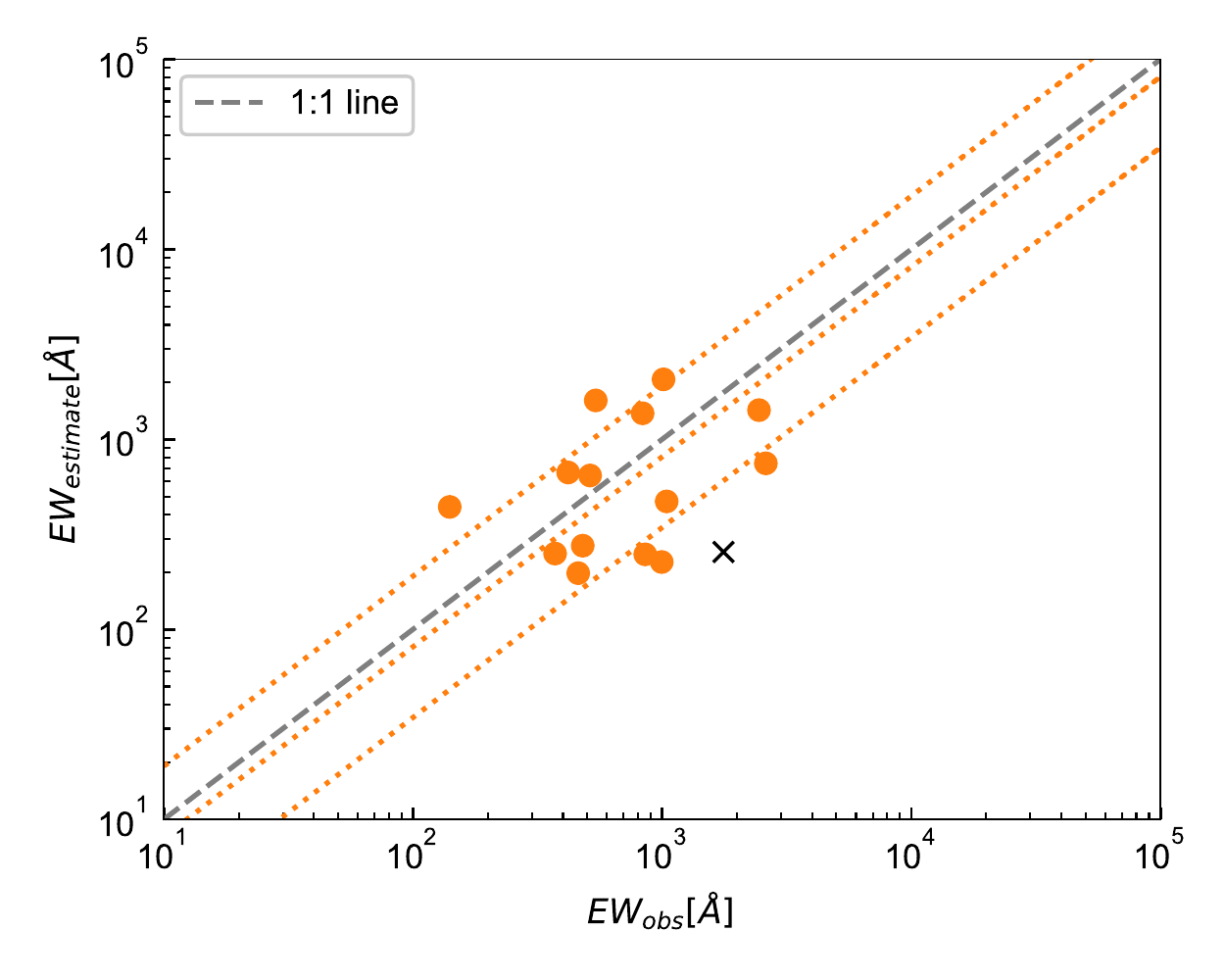}
\caption{Comparison of $K_s$-band excess and [OIII]+H$\beta$ fluxes for Sample-A galaxies. The crosses show the two catastropic cases. The three dotted lines indicate the mean and standard deviations of the $\Delta$EW, where $\Delta$EW=$\log$EW$_\mathrm{obs}$-$\log$EW$_\mathrm{model}$. \label{fig:kbandex}}
\end{figure}

As mentioned in Section \ref{subsec:compprop}, although the Withline and Noline fits output similar $M_*$, SFR, and $A_V$ for our sample, the Withline fit has more constraints on the outputs since it includes more information provided by the line emission data. We find that the uncertainties of the parameters are reduced by including the emission line data. On average, the uncertainties are reduced by 0.05 dex, 0.06 dex, and 0.05 dex for $M*$, SFR, and $A_V$, respectively. For some galaxies ($\sim$ 30\% of the sample), the reduction is larger than 0.1 dex. Figure \ref{fig:constraint} shows an example of the pdf of the output parameters. From the comparison of the pdf derived from the Withline and Noline fits, we can see clearly that the inclusion of the emission line data improves the constraints on these parameters.

\subsection{Improvement of the line flux estimation}\label{subsec:lineflux}
Besides of the output parameters discussed in \ref{subsec:compprop}, we also find that without emission line data, the Noline fit cannot predict line fluxes consistent with the observation. Figure \ref{fig:linecomp_woline} shows the comparison of the line fluxes obtained by the Withline and Noline fit with the observation. The line fluxes obtained by the Withline fit are consistent with the observation, while those obtained by the Noline fit are not, implying the importance of including the emission line observation data in the fitting procedure.

\subsection{Reliability of the line fluxes estimated from the $K_s$-band excess}\label{subsec:linepredict}
\citet{stark2013} found that the broadband flux excesses can be used to infer line strengths in carefully selected spectroscopic samples at higher redshifts where direct spectroscopic measurements of nebular line fluxes are not yet available. They examined the SEDs of a sample of nebular line emitters at $1.3<z<2.3$ with spectroscopically measured [OIII] line fluxes from WFC3 grism observations of the Hubble Ultra Deep Field (UDF; \citealt{trump2011}). By comparing the observed broadband flux (in the contaminated filter, $J_{125}$ or $H_{160}$)) to the stellar continuum flux expected from the best-fitting population synthesis models, they characterized the strength of the emission lines and found that the results were consistent with those measured spectroscopically. Based on this method, the $K_s$-band excess is used to indicate the [OIII] fluxes for galaxies at $z \sim 3$ \citep[e.g.,][] {forrest2017, malkan2017}. However, since the direct comparison of this estimation to the spectroscopic data made by \citet{stark2013} is only for galaxies at $z<2.3$, with the photometric and spectroscopic data of our galaxies, we further examine the reliability of this estimation for galaxies at higher redshifts ($z\sim3.5$).

Taking advantage of the spectroscopic data of our sample, we compare the $K_s$-band excess with the observed f([OIII]+H$\beta$) to test this method. We use only galaxies in Sample-A for this test because they have more data points to constrain the fluxes. The $K_s$-band excess is calculated by $f_\mathrm{K,obs}-f_\mathrm{K,mod}$, where $f_\mathrm{K,obs}$ is the observed fluxes, and $f_\mathrm{K,mod}$ is the model fluxes derived from the Nonebular fit. As described in Section \ref{sec:method}, the Nonebular fit is a simple stellar population synthesis, which is consistent with the models that have been used in other works \citep[e.g.,][]{stark2013,malkan2017}. The excess of the observed flux to the continuum flux expected from the model in $K_s$-band is then interpreted as the [OIII]+H$\beta$~line contamination. The results are shown in Figure \ref{fig:kbandex}.

Figure \ref{fig:kbandex} shows that the EWs estimated from the $K_s$-band excess correlates well with the observations except for one catastrophic cases (marked as the cross in the figure). We check the fitting results of these two galaxies and find that the $\chi^2$ values are too small ($\sim 0.2$), meaning that the data for these two galaxies have large uncertainties so that the fit cannot detect the excess due to the emission lines.

Excluding this case, we find that the mean EW$_\mathrm{model}$ is $759$\AA, and the mean EW$_\mathrm{obs}$ is $908$\AA. Both have large scatters ($\sim711$\AA). We calculate the difference $\Delta$EW between $\log \mathrm{EW_{obs}}$ and $\log \mathrm{EW_{model}}$, and find that the mean $\Delta$EW is $-0.09$ dex, with a standard deviation of $0.37$ dex. The results confirm that the estimation of line fluxes from the $K_s$-band excess is statistically consistent with the observation. However, there are large uncertainties in the estimation for individual galaxies.

In the above analysis, we include the contaminated filters in the fitting process. We further examine the stellar models excluding the contaminated bandpass and find that the inferred line flux is typically two times greater than the spectroscopic measurement. This result agrees with what has been found by \citet{stark2013}. As explained by \citet{stark2013}, since the SEDs are quite poorly sampled in the wavelength range around the D4000 break, without the contaminated filter, the fitting process will take prior to fit the adjacent filters but underpredict the D4000 break and thus the continuum in the vicinity of the emission line of interest.

\subsection{Balmer absorption \label{subsec:absorption}}

In the works of S13, T14, and H16, the fluxes of emission lines are calculated without considering the underline absorption of the stellar continuum. The continuum is subtracted simply using linear interpolation. However, for Balmer lines, such as H$\beta$, the absorption from stellar continuum can be strong, and thus the measured line fluxes can be quite different from the flux of the nebular model because the measurement is the sum of the absorption and the emission.

 From the best model of the SED fitting, we can estimate the contribution of the stellar absorption to the measured line fluxes. From the best fitting model of each galaxy, we calculate the EW of the H$\beta$ absorption, EW$_{\mathrm{H}\beta}^-$, from the stellar model. Then we calculate the EW of the H$\beta$ emission, EW$_{\mathrm{H}\beta}^+$, from the nebular model. In both calculations, the continuum is estimated from the mean of 4 nm spectrum each on the blue and red side of the H$\beta$ line from the best fitting model spectrum.

The mean EW$_{\mathrm{H}\beta}^-$ of our sample, $<$EW$_{\mathrm{H}\beta}^->$, is 4.4{\AA} for our sample in the observational frame, and $\sim$1{\AA} in the rest-frame. This result is smaller than the estimation given in Brinchmann et al. (2004), where the $<$EW$_{\mathrm{H}\beta}^->$ is estimated as {$\sim$3.3{\AA} (assuming $<$EW$_{\mathrm{H}\alpha}^->$=$0.6<$EW$_{\mathrm{H}\beta}^->$)}. The smaller EW$_{\mathrm{H}\beta}^-$ means that the stellar populations in these galaxies are so young that the absorption features produced by older stellar populations are not significant. The mean EW of the H$\beta$ emission $<$EW$_{\mathrm{H}\beta}^+>$ is 236.3{\AA} in the observational frame, and 56.3{\AA} in the rest-frame. We also calculated the ratio of the absorption to the emission, $Q\equiv \mathrm{EW}_{\mathrm{H}\beta}^-/\mathrm{EW}_{\mathrm{H}\beta}^+$, and find that the average is about 0.04, indicating that the absorption from the stellar continuum is insignificant in the study of the line fluxes.

\section{Discussion \label{sec:discussion}}

\begin{figure}
\centering
\includegraphics[width=\linewidth]{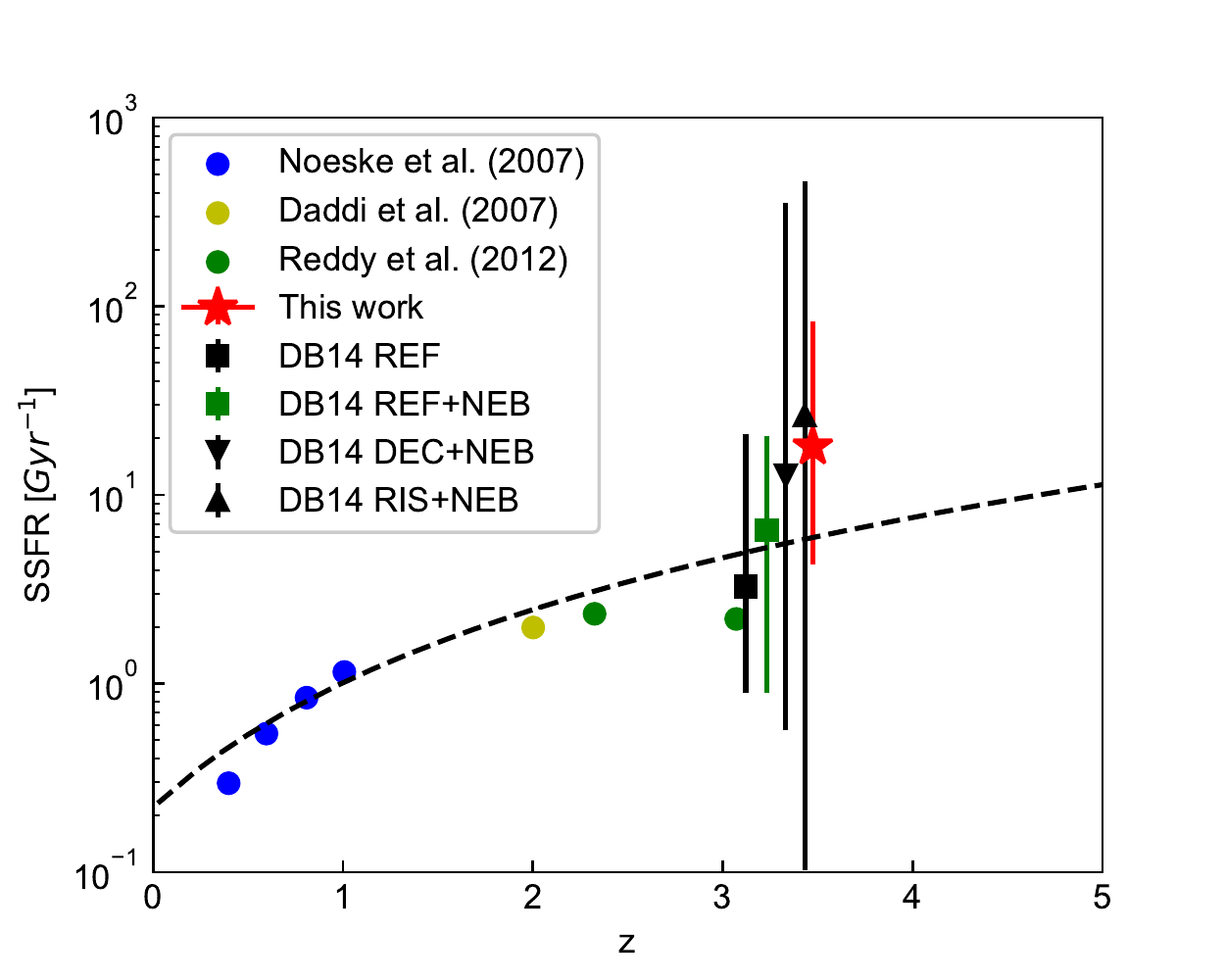}
\caption{Median SSFR as a function of redshift with a 68\% confidence limit based on the whole PDF with comparison of results from different studies \citep{noeske2007,daddi2007, reddy2012} and results from the study of \citet{debarros2014} (DB14) for four different models. The dashed line shows the relation expected from \citet{bouche2010} for an exponentially increasing star formation at a fixed stellar mass, $10^{9.5}M_{\odot}$. \label{fig:ssfr_z}}
\end{figure}

\begin{figure}
\centering
\includegraphics[width=\linewidth]{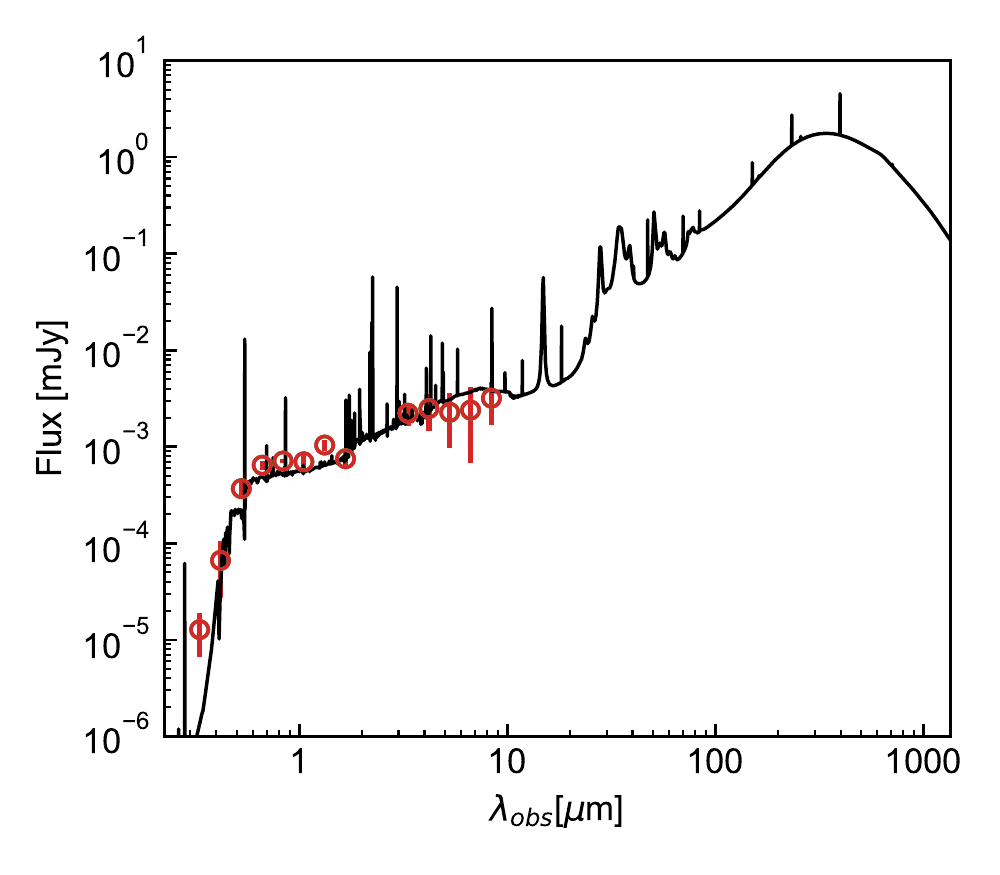}
\caption{SEDs of LBG galaxies shifted to $z=3.5$ and normalized to the flux at rest 0.2$\mu$m (0.9 $\mu$m in this figure). Red circles show the composite SED with the errorbars corresponding to the dispersion of the data. The normalization of the composite SED at 0.9 $\mu$m is 0.71 $\mu$Jy. The line is the SED templates given by \citet{avarez2019}. \label{fig:stack}}
\end{figure}

\subsection{Properties of LBGs with detected emission lines at $z\sim 3.5$}

The LBGs in our sample have detected emission lines, indicating that they are more actively star forming than LBGs with no or weak emission lines. To examine how representative our galaxies are for typical $z\sim3.5$ galaxies, we compare the average specific star formation rate (SSFR) of our galaxies with those derived from other studies, as shown in Figure \ref{fig:ssfr_z}. We use the compilation from \citet{gonzalez2010} and \citet{debarros2014}, including the results from \citet{noeske2007}, \citet{daddi2007}, \citet{reddy2012}, and \citet{debarros2014}. Among them, \citet{debarros2014} provide results from different models. The REF, DEC and RIS models assume constant SFH, exponentially declined SFH, and rising SFH, respectively. The ``+NEB'' model includes the nebular emission.
The median mass of our galaxies is $10^{9.7}M_{\odot}$, which is close to the results given by the REF model in \citet{debarros2014}.

We find that the median SSFR of our galaxy sample is higher than the median SSFRs estimated in \citet{reddy2012} and \citet{debarros2014} except for the RIS+NEB model. The higher median SSFR can be explained by the fact that the galaxies with emission lines are more actively forming stars than those with weak or no emission lines, as concluded by \citet{debarros2014}. However, the comparison is also dependent on the model used in the SED fitting. Compared to the result using a rising SFH (RIS+NEB) in \citet{debarros2014}, our galaxies have lower median SSFR. Moreover, both samples in the work of \citet{debarros2014} and this work are incomplete. Due to the incompleteness, the results can be considered as a lower limit {considering the trend of increasing SSFR with decreasing stellar mass \citep{debarros2014}}. A more detailed study on the SSFRs for galaxies at this redshift range requires a more statistically complete sample.

\subsection{Composite SED}
 We further investigate the properties of our galaxies with the stacking method. We aim to compare the composite SED of our galaxies with the template SED given by \citet{avarez2019}, who selected more than 10,000 LBGs in the COSMOS field and obtained empirical UV-to-IR SED templates of LBG templates at $z\sim3.5$ by stacking the images at multiple wavelengths. For this comparison, we use galaxies with $\log M_{*}>9.75$ (24 galaxies) because in \citet{avarez2019}, the mass range is from $\log M_{*}>9.75$. We check the distribution of the stellar mass of these 24 galaxies and find that the shape of the distribution is similar to that in \citet{avarez2019}.

 We adopt the same method described in \citet{hirashita2017} to stack the SEDs of these galaxies. First, we redshift the SEDs of these galaxies into the same redshift ($z=3.5$). Second, we normalize the SED of each galaxy to the flux at rest 0.2 $\mu$m (i.e., observed wavelength is 0.9 $\mu$m), in order to extract the information on the SED shape. Then we divide the data into 15 bins with a logarithmically equal width. We take the median values of the normalized data points in each bin to obtain the composite SED. Finally, the normalized flux is multiplied by the logarithmically averaged rest 0.2 $\mu$m (0.71 $\mu$Jy) to obtain the absolute level of the flux. The composite SED is shown in Figure \ref{fig:stack}.

 The template SED given by \citet{avarez2019} is plotted in Figure \ref{fig:stack} for comparison (re-scaled to the same stellar mass as our composite SED). We find that the composite SED of our galaxies with [OIII] detections has a similar shape to galaxies in \citet{avarez2019}. However, we can see that our composite SED is slightly bluer, {which may suggest that the galaxies with emission lines have a higher sSFR, or they are with less dust}. Further investigation of the dust attenuation properties of these galaxies requires data from far-IR.

\subsection{Constraints on the models \label{subsec:pardiscussion}}

We examine whether the photometric and line data used in this work could put some constraints on different star formation and dust attenuation models. First, we compare four different SFH models, including the delayed SFH (used in the above sections), exponentially decreasing SFH, exponentially rising SFH, and the delay+burst SFH. {The functional forms of these models are listed in Table \ref{tab:sfh}. In these models, $t$ is the stellar age from the onset of the star formation, and $\tau$ is the star formation time-scale. The delay+burst model is more complicated than the other three models. It allows for an second exponential burst representing the latest episode of star formation \citep{malek2018,boquien2019}. In the model, $t_0$ and $t_1$ are the time of the onset of the first and the second episodes of star formation, respectively, $\tau_1$ is the e-folding time of the burst, and $k$ the relative amplitude of the burst component.}

\begin{table*}
    \centering
    \caption{SFH models used for comparison.}
    \begin{tabular}{l l l}
    \hline
    SFH Model     &  SFR($t$) & $<\chi^2_r>$ \\
    \hline
    Delayed     &  $\propto t/\tau^2 \exp{(-t/\tau)}$  &  1.25 \\
    Exponentially declining & $\propto \exp{(-t/\tau)}$ & 1.31\\
    Exponentially rising & $\propto \exp{(t/\tau)}$  & 1.32 \\
    \multirow{2}{*}{Delay+burst}  &  $\propto t/\tau^2 \exp{(-t/\tau)}$ if $t<t_0-t_1$\\
                 &  $\propto t/\tau^2 \exp{(-t/\tau)}+k\times \exp{(-t/\tau_1)}$ if $t>t_0-t_1$ &  1.43\\
    \hline
    \end{tabular}

    \label{tab:sfh}
\end{table*}

The comparison shows that the four models result in similar $\chi^2$ values (see Table \ref{tab:sfh}). Furthermore, we compare the Bayesian information criterion (BIC) which accounts for the increase of free parameters. We calculated the BIC values using a simplified form, BIC$=\chi^2+k\times \ln{(n)}$, where $\chi^2$ is the non-reduced of the best-fit model, $k$ the number of free parameters, and $n$ the number of data fit \citep{buat2018,ciesla2018}. The BIC values for these four different models are quite similar. For most galaxies ($\sim 85\%$), the difference between BIC(delayed SFH) and BIC(rising or decreasing SFH) is smaller than 2, meaning that the evidence against the model with the higher BIC is not worth more than a bare mention. The difference between BIC(delay+burst) and BIC(delay) is higher than 2 for {80\%} of our galaxies, indicating positive evidence against the delay+burst model, which has two more free parameters than the other models. However, the difference is larger than 6 for only nine galaxies, i.e., the evidence is not strong for most galaxies in our sample. The BIC analysis for these models suggests that our data are not sufficient to discriminate these SFHs.

We also compare different attenuation curves. We compare the Calzetti Law and the modified Calzetti Law and find that the average $\chi^2_r$ for the fit using the Calzetti law is $1.52$, about $\sim 20\%$ larger than that with the modified Calzetti law. Considering the different number of the input parameters, we compare the values of the BIC. The difference between BIC(Calzetti) and BIC(modified-Calzetti) is found higher than 6 (corresponding to a strong evidence against the model with the higher BIC) only for three galaxies ($\sim$6\% of the sample), meaning that the photometric and line data used in our work cannot discriminate these attenuation curves for most of our galaxies.

Providing additional data may improve the constraints on the models. First, it is essential to obtain H$\alpha$ data for these galaxies. The [OIII] and [OII] data can provide some constraints on the nebular models, but they depend strongly on the metallicity, and therefore cannot entirely break the degeneracy between the metallicity and the dust attenuation. As shown in \citet{buat2018}, the H$\alpha$ data have helped constrain the dust attenuation curves. With H$\alpha$ and H$\beta$ data, the derived Balmer decrement can add a strong constraint on the dust attenuation in the ISM. Second, restframe mid-to-far IR data can also help improve the constraints on the dust attenuation. We note that using different models does not affect our conclusion obtained in Section \ref{sec:result}.

\section{Summary and conclusion \label{sec:conclusion}}
We collected a sample of LBGs with emission line data at $z\sim3.5$. To investigate the impact of including emission line data in the SED fitting, we compare three different fits. The Withline fit uses the nebular and stellar population models on both the line and photometry data. The Noline fit uses the nebular and stellar population models, but only fit the photometry data. The Nonebular fit does not consider the nebular model. It fit the photometry data using the stellar population model only. By comparing the output of these fits, we conclude as follows:

The Nonebular fit overestimates the old stellar populations and thus overestimates the stellar masses because it does not take into account the contribution of the emission lines (such as [OIII] and H$\beta$) to the broadband fluxes. The Noline fit can give similar results of $M_*$, SFR and $A_V$ compared to the Withline fit. However, in the Noline fit the uncertainties of these parameters are considerably larger than in the Withline fit. Also, without the emission line data, the Noline fit cannot predict the line fluxes reliably.

By comparing the line fluxes estimated from models to the observed line fluxes, we find that it is risky to predict line fluxes using photometric data alone. The method of the $K_s$-band excess is statistically right, but it may cause significant errors if used for individual galaxies. We also find that Balmer absorption plays an insignificant role in measuring the line fluxes, consistent with the fact that the stellar populations in these galaxies are not old enough to produce deep absorption features.

Our UV-to-NIR SED and the emission line data ([OII], [OIII], H$\beta$, etc.) together still cannot constrain dust attenuation law and SFH very well. The H$\alpha$ emission line data provided by future observations (e.g., JWST) may help anchor dust attenuation and thus better constrain the properties of these galaxies.

\begin{acknowledgement}

We thank the anonymous referee for helpful and constructive comments that improve the paper. This work is partly supported by the National Natural Science Foundation of China (NSFC) under grant nos.11433003, 11573050, and by a China-Chile joint grant from CASSACA. FTY acknowledges support from the Natural Science Foundation of Shanghai (Project Number: 17ZR1435900).
MB acknowledges the FONDECYT regular grant 1170618.
This work is based on observations taken by the 3D-HST Treasury Program (GO 12177 and 12328) with the NASA/ESA HST, which is operated by the Association of Universities for Research in Astronomy, Inc., under NASA contract NAS5-26555.
\end{acknowledgement}

\bibliography{ref}
\bibliographystyle{aa}




\end{document}